\documentstyle[12pt,epsfig]{article}
\begin{document}

\baselineskip18pt

\begin{center}
{\Large\bf QUANTUM DECOHERENCE \\IN THE THEORY OF OPEN SYSTEMS}

\vspace{1cm}

{\it Aurelian Isar $^a$}

\vspace{0.5cm}

$^a$ Department of Theoretical Physics,
Institute of Physics and Nuclear Engineering\\
Bucharest-Magurele, Romania\\
e-mail: isar@theory.nipne.ro
\end{center}

\vspace{1cm}

\begin{abstract}
In the framework of the Lindblad theory for open quantum systems, we
determine the degree of quantum decoherence of a harmonic oscillator
interacting with a thermal bath. It is found that the system
manifests a quantum decoherence which is more and more significant
in time. We calculate also the decoherence time scale and analyze
the transition from quantum to classical behaviour of the considered
system.
\end{abstract}

\section{Introduction}

The quantum to classical transition and classicality of quantum
systems continue to be among the most interesting problems in many
fields of physics, for both conceptual and experimental reasons
\cite{1,2}. Two conditions are essential for the classicality of a
quantum system \cite{3}: a) quantum decoherence (QD), that means the
irreversible, uncontrollable and persistent formation of a quantum
correlation (entanglement) of the system with its environment
\cite{4}, expressed by the damping of the coherences present in the
quantum state of the system, when the off-diagonal elements of the
density matrix decay below a certain level, so that this density
matrix becomes approximately diagonal and b) classical correlations,
expressed by the fact that the quantum state becomes peaked along a
classical trajectory. Classicality is an emergent property of open
quantum systems, since both main features of this process -- QD and
classical correlations -- strongly depend on the interaction between
the system and its external environment \cite{1,2}. In this work we
study QD and analyze quantum-classical transition of a harmonic
oscillator interacting with an environment, in particular with a
thermal bath, in the framework of the Lindblad theory for open
quantum systems.

\section{Master equation and density matrix}

In the Lindblad axiomatic formalism based on quantum dynamical
semigroups, the irreversible time evolution of an open system is
described by the following general quantum Markovian master equation
for the density operator $\rho(t)$ \cite{5}:
\begin{eqnarray}{d \rho(t)\over dt}=-{i\over\hbar}[H,\rho(t)]
+{1\over 2\hbar} \sum_{j}([  V_{j} \rho(t), V_{j}^\dagger ]+[ V_{j},
\rho(t) V_{j}^\dagger ]).\label{lineq}\end{eqnarray} The harmonic
oscillator Hamiltonian $H$ is chosen of the general quadratic form
\begin{eqnarray} H=H_{0}+{\mu\over 2}(qp+pq), ~~~  H_{0}={1\over
2m}p^2+{m\omega^2\over 2}  q^2 \label{ham} \end{eqnarray} and the
operators $V_{j},$ $ V_{j}^\dagger,$ which model the environment,
are taken as linear polynomials in coordinate $q$ and momentum $p.$
Then the master equation (\ref{lineq}) takes the following form
\cite{6}:
\begin{eqnarray} {d \rho \over dt}=-{i\over \hbar}[ H_{0}, \rho]-
{i\over 2\hbar}(\lambda +\mu) [q, \rho p+ p \rho]+{i\over
2\hbar}(\lambda -\mu)[  p,
\rho   q+  q \rho]  \nonumber\\
-{D_{pp}\over {\hbar}^2}[  q,[  q, \rho]]-{D_{qq}\over {\hbar}^2} [
p,[  p, \rho]]+{D_{pq}\over {\hbar}^2}([  q,[  p, \rho]]+ [ p,[ q,
\rho]]). ~~~~\label{mast}   \end{eqnarray} The diffusion
coefficients $D_{pp},D_{qq},$ $D_{pq}$ and the dissipation constant
$\lambda$ satisfy the fundamental constraints: $ D_{pp}>0, D_{qq}>0$
and $D_{pp}D_{qq}-D_{pq}^2\ge {\lambda}^2{\hbar}^2/4.$ In the
particular case when the asymptotic state is a Gibbs state $
\rho_G(\infty)=e^{-{  H_0\over kT}}/ {\rm Tr}e^{-{ H_0\over kT}}, $
these coefficients become
\begin{eqnarray} D_{pp}={\lambda+\mu\over 2}\hbar
m\omega\coth{\hbar\omega\over 2kT}, ~~D_{qq}={\lambda-\mu\over
2}{\hbar\over m\omega}\coth{\hbar\omega\over 2kT}, ~~D_{pq}=0,
\label{coegib}
\end{eqnarray} where $T$ is the temperature of the thermal bath. In
this case, the fundamental constraints are satisfied only if
$\lambda>\mu$ and
\begin{eqnarray} (\lambda^2-\mu^2)\coth^2{\hbar\omega\over 2kT}
\ge\lambda^2\label{cons}\end{eqnarray} and the asymptotic values
$\sigma_{qq}(\infty),$ $\sigma_{pp}(\infty),$ $\sigma_{pq}(\infty)$
of the dispersion (variance), respectively correlation (covariance),
of the coordinate and momentum, reduce to \cite{6}
\begin{eqnarray} \sigma_{qq}(\infty)={\hbar\over
2m\omega}\coth{\hbar\omega\over 2kT}, ~~\sigma_{pp}(\infty)={\hbar
m\omega\over 2}\coth{\hbar\omega\over 2kT}, ~~\sigma_{pq}(\infty)=0.
\label{varinf} \end{eqnarray}

We consider a harmonic oscillator with an initial Gaussian wave
function ($\sigma_q(0)$ and $\sigma_p(0)$ are the initial averaged
position and momentum of the wave packet) \begin{eqnarray}
\Psi(q)=({1\over 2\pi\sigma_{qq}(0)})^{1\over 4}\exp[-{1\over
4\sigma_{qq}(0)}
(1-{2i\over\hbar}\sigma_{pq}(0))(q-\sigma_q(0))^2+{i\over
\hbar}\sigma_p(0)q], \label{ccs}\end{eqnarray} representing a
correlated coherent state (squeezed coherent states) with the
variances and covariance of coordinate and momentum
\begin{eqnarray} \sigma_{qq}(0)={\hbar\delta\over 2m\omega},~~
\sigma_{pp}(0)={\hbar m\omega\over 2\delta(1-r^2)},~~
\sigma_{pq}(0)={\hbar r\over 2\sqrt{1-r^2}}.
\label{inw}\end{eqnarray} $\delta$ is the squeezing parameter which
measures the spread in the initial Gaussian packet and $r,$ with
$|r|<1$ is the correlation coefficient. The initial values
(\ref{inw}) correspond to a minimum uncertainty state, since they
fulfil the generalized uncertainty relation $
\sigma_{qq}(0)\sigma_{pp}(0)-\sigma_{pq}^2(0) =\hbar^2/4.$ For
$\delta=1$ and $r=0$ the correlated coherent state becomes a Glauber
coherent state.

From Eq. (\ref{mast}) we derive the evolution equation in coordinate
representation: \begin{eqnarray} {\partial\rho\over\partial
t}={i\hbar\over 2m}({\partial^2\over\partial q^2}-
{\partial^2\over\partial q'^2})\rho-{im\omega^2\over
2\hbar}(q^2-q'^2)\rho\nonumber\\
-{1\over 2}(\lambda+\mu)(q-q')({\partial\over\partial
q}-{\partial\over\partial q'})\rho+{1\over
2}(\lambda-\mu)[(q+q')({\partial\over\partial
q}+{\partial\over\partial
q'})+2]\rho  \nonumber\\
-{D_{pp}\over\hbar^2}(q-q')^2\rho+D_{qq}({\partial\over\partial
q}+{\partial\over \partial q'})^2\rho -{2iD_{pq}\hbar}(q-q')(
{\partial\over\partial q}+{\partial\over\partial
q'})\rho.\label{cooreq}\end{eqnarray} The first two terms on the
right-hand side of this equation generate the usual Liouvillian
unitary evolution. The third and forth terms are the dissipative
terms and have a damping effect (exchange of energy with
environment). The last three are noise (diffusive) terms and produce
fluctuation effects in the evolution of the system. $D_{pp}$
promotes diffusion in momentum and generates decoherence in
coordinate $q$ -- it reduces the off-diagonal terms, responsible for
correlations between spatially separated pieces of the wave packet.
Similarly $D_{qq}$ promotes diffusion in coordinate and generates
decoherence in momentum $p.$ The $D_{pq}$ term is the so-called
"anomalous diffusion" term and it does not generate decoherence.

The density matrix solution of Eq. (\ref{cooreq}) has the general
Gaussian form \begin{eqnarray} <q|\rho(t)|q'>=({1\over
2\pi\sigma_{qq}(t)})^{1\over 2} \exp[-{1\over
2\sigma_{qq}(t)}({q+q'\over
2}-\sigma_q(t))^2\nonumber\\
-{\sigma(t)\over 2\hbar^2\sigma_{qq}(t)}(q-q')^2
+{i\sigma_{pq}(t)\over \hbar\sigma_{qq}(t)}({q+q'\over
2}-\sigma_q(t))(q-q')+{i\over
\hbar}\sigma_p(t)(q-q')],\label{densol} \end{eqnarray} where
$\sigma(t)\equiv\sigma_{qq}(t)\sigma_{pp}(t)-\sigma_{pq}^2(t)$ is
the Schr\"odinger generalized uncertainty function. In the case of a
thermal bath we obtain the following steady state solution for
$t\to\infty$ ($\epsilon\equiv{\hbar\omega\over 2kT}$):
\begin{eqnarray} <q|\rho(\infty)|q'>=({m\omega\over
\pi\hbar\coth\epsilon})^{1\over 2}\exp\{-{m\omega\over
4\hbar}[{(q+q')^2\over\coth\epsilon}+
(q-q')^2\coth\epsilon]\}.\label{dinf}\end{eqnarray}

\section{Decoherence and quantum-classical transition}

An isolated system has an unitary evolution and the coherence of the
state is not lost -- pure states evolve in time only to pure states.
The QD phenomenon, that is the loss of coherence or the destruction
of off-diagonal elements representing coherences between quantum
states in the density matrix, can be achieved by introducing an
interaction between the system and environment: an initial pure
state with a density matrix which contains nonzero off-diagonal
terms can non-unitarily evolve into a final mixed state with a
diagonal density matrix.

Using new variables $\Sigma=(q+q')/2$ and $\Delta=q-q',$ the density
matrix (\ref{densol}) becomes \begin{eqnarray}
\rho(\Sigma,\Delta,t)=\sqrt{\alpha\over \pi}\exp[-\alpha\Sigma^2
-\gamma\Delta^2
+i\beta\Sigma\Delta+2\alpha\sigma_q(t)\Sigma+i({\sigma_p(t)\over\hbar}-
\beta\sigma_q(t))\Delta-\alpha\sigma_q^2(t)],\label{ccd3}\end{eqnarray}
with the abbreviations \begin{eqnarray} \alpha={1\over
2\sigma_{qq}(t)},~~\gamma={\sigma(t)\over 2\hbar^2
\sigma_{qq}(t)},~~ \beta={\sigma_{pq}(t)\over\hbar\sigma_{qq}(t)}.
\label{ccd4}\end{eqnarray}

The representation-independent measure of the degree of QD \cite{3}
is given by the ratio of the dispersion $1/\sqrt{2\gamma}$ of the
off-diagonal element $\rho(0,\Delta,t)$ to the dispersion
$\sqrt{2/\alpha}$ of the diagonal element $\rho(\Sigma,0,t):$
\begin{eqnarray} \delta_{QD}(t)={1\over 2}\sqrt{\alpha\over
\gamma}={\hbar\over 2\sqrt{\sigma(t)}}.\label{qdec}\end{eqnarray}

The finite temperature Schr\"odinger generalized uncertainty
function has the expression \cite{7} (with the notation
$\Omega^2\equiv\omega^2-\mu^2$,
$\omega>\mu$)\begin{eqnarray}\sigma(t)={\hbar^2\over
4}\{e^{-4\lambda
t}[1-(\delta+{1\over\delta(1-r^2)})\coth\epsilon+\coth^2\epsilon]\nonumber\\
+e^{-2\lambda t}\coth\epsilon[(\delta+{1\over\delta(1-r^2)}
-2\coth\epsilon){\omega^2-\mu^2\cos(2\Omega
t)\over\Omega^2}\nonumber\\ +(\delta-{1\over\delta(1-r^2)}){\mu
\sin(2\Omega t)\over\Omega}+{2r\mu\omega(1-\cos(2\Omega
t))\over\Omega^2\sqrt{1-r^2}}]+\coth^2\epsilon\}.\label{sunc}\end{eqnarray}
In the limit of long times Eq. (\ref{sunc}) yields
$\sigma(\infty)=(\hbar^2\coth^2\epsilon)/4,$ so that we obtain
\begin{eqnarray} \delta_{QD}(\infty)=\tanh{\hbar\omega\over
2kT},\end{eqnarray} which for high $T$ becomes
$\delta_{QD}(\infty)=\hbar\omega/(2kT).$ We see that $\delta_{QD}$
decreases, and therefore QD increases, with time and temperature,
i.e. the density matrix becomes more and more diagonal at higher $T$
and the contributions of the off-diagonal elements get smaller and
smaller. At the same time the degree of purity decreases and the
degree of mixedness increases with $T.$ For $T=0$ the asymptotic
(final) state is pure and $\delta_{QD}$ reaches its initial maximum
value 1. $\delta_{QD}= 0$ when the quantum coherence is completely
lost, and if $\delta_{QD}= 1$ there is no QD. For long enough time
the magnitude of the elements of the density matrix in the position
basis are peaked preferentially along the diagonal $q=q'.$ Then
$\delta_{QD}<1$ and we can say that the considered system
interacting with the thermal bath manifests QD. Dissipation promotes
quantum coherences, whereas fluctuation (diffusion) reduces
coherences and promotes QD. The balance of dissipation and
fluctuation determines the final equilibrium value of $\delta_{QD}.$
The initial pure state evolves approximately following the classical
trajectory in phase space and becomes a quantum mixed state during
the irreversible process of QD.

In the macroscopic limit, when $\hbar$ is small compared to other
quantities with dimensions of action, the term in Eq. (\ref{cooreq})
containing $D_{pp}/\hbar^2$ dominates and induces the following
evolution:
\begin{eqnarray} {\partial\rho\over\partial t}=
-{D_{pp}\over\hbar^2}(q-q')^2\rho.\label{cooreq1}\end{eqnarray} Thus
the density matrix loses off-diagonal terms in position
representation, while the diagonal ($q=q'$) ones remain untouched.
Quantum coherences decay exponentially and the decoherence time
scale is of the order of \begin{eqnarray}
t_{deco}={\hbar^2\over{D_{pp}(q-q')^2}}.\end{eqnarray} In the case
of a thermal bath, we obtain (see Eq. (\ref{coegib}))
\begin{eqnarray} t_{deco}={2\hbar\over{(\lambda+\mu)m\omega
\sigma_{qq}(0)\coth \epsilon}},\label{tdeco}\end{eqnarray} where we
have taken $(q-q')^2$ of the order of the initial dispersion in
coordinate $\sigma_{qq}(0).$ As expected, the decoherence time
$t_{deco}$ has the same scale as the time after which thermal
fluctuations become comparable with quantum fluctuations. In the
macroscopic domain QD occurs very much faster than relaxation. When
$t\gg t_{rel},$ where $t_{rel}\approx\lambda^{-1}$ is the relaxation
time, which governs the rate of dissipation, the particle reaches
equilibrium with the environment. Indeed, the uncertainty function
$\sigma(t)$ (\ref{sunc}) approaches
$\sigma^{BE}=(\hbar^2\coth^2\epsilon)/4,$ which is the Bose-Einstein
relation for a system of bosons in equilibrium at temperature $T.$
In the case of $T=0$ we approach the limit of pure quantum
fluctuations, $\sigma_0=\hbar^2/4,$ which is the quantum Heisenberg
relation. At high temperatures $T$ we obtain the limit of pure
thermal fluctuations, $\sigma^{MB}=({kT/\omega})^2,$ which is a
Maxwell-Boltzmann distribution for a system approaching a classical
limit. The regime where thermal fluctuations begin to surpass
quantum fluctuations is regarded as the transition point from
quantum to classical statistical mechanics and the high temperature
regime of a system is considered as the classical regime. We have
shown that these two criteria of classicality are equivalent: the
time when the quantum system decoheres is comparable with the time
when thermal fluctuations overtake quantum fluctuations. After the
decoherence time, the system has to be described by non-equilibrium
quantum statistical mechanics. After the relaxation time the system
is treated by equilibrium quantum statistical mechanics, and only at
a sufficiently high temperature, when the spin statistics is
represented by the Maxwell-Boltzmann distribution function, it can
be considered in a classical regime \cite{7,8}.

\section{Summary and concluding remarks}

We have studied QD in the framework of the theory of open quantum
systems in order to understand the quantum-classical transition for
a harmonic oscillator in interaction with a thermal bath. The
classicality is conditioned by QD, expressed by the loss of quantum
coherence in the case of a thermal bath at finite temperature.

The role of QD became relevant in many interesting physical
problems. In many cases one is interested in understanding QD
because one wants to prevent decoherence from damaging quantum
states and to protect the information stored in quantum states from
the degrading effect of the interaction with the environment.
Decoherence is also responsible for washing out the quantum
interference effects which are desirable to be seen as signals in
experiments. QD has a negative influence on many areas relying upon
quantum coherence effects, in particular it is a major problem in
the physics of quantum information and computation.

\section*{Acknowledgments}

The author acknowledges the financial support received within the
"Hulubei-Meshcheryakov" Programme, JINR order No. 328/20.05.2005.

\end{document}